\documentclass{PoS}

\title{End-to-end data acquisition pipeline for the Cherenkov Telescope Array}

\ShortTitle{Data acquisition pipeline for CTA}

\author{\speaker{E. Lyard}, R. Walter for the CTA Consortium\\
        ISDC, University of Geneva, Switzerland\\
        E-mail: \email{etienne.lyard@unige.ch roland.walter@unige.ch}}

\abstract{The Cherenkov Telescope Array (CTA) will operate several types of telescopes and cameras.
		  The individual camera trigger rates will vary much - from 0.6 to 15 kHz - while the content 
		  of the raw data will be heterogeneous. Raw data streams of up to 43 Gbps per telescope must 
	      be handled efficiently, from the camera front-ends down to the on-site repository and 
		  real-time analysis. In addition, the system must transcode all raw data to a common, 
		  pre-calibrated format. \\
		  \
		  We will present the pipeline that we propose to implement this data acquisition pipeline. 
		  It will format the raw data to a common structure, provide facilities to run camera-specific 
		  algorithms and compress and write data to the on-site repository. We will also present the 
		  Python interface that allows the analysis pipeline to access the data. Eventually, the two 
	      strategies foreseen to interface the camera servers will be detailed and the current status
		  of the developments for CTA will be given, with the last performance figures measured.}

\FullConference{35th International Cosmic Ray Conference --- ICRC2017\\
		10--20 July, 2017\\
		Bexco, Busan, Korea}

\begin{document}

\section{Introduction}
The Cherenkov Telescope Array (CTA) \cite{cta} will operate more than 100 telescopes of different sizes between 4 and 23 meters in diameter. 
Each type of telescope will have one or two different kinds of Cherenkov camera, leading to very heterogenous data outputted by 
the telescopes. This document presents the pipeline that we propose to deploy to handle these data along with performance figures 
obtained from the prototype implementation of it. \\

\begin{figure}
\includegraphics[width=.9\textwidth]{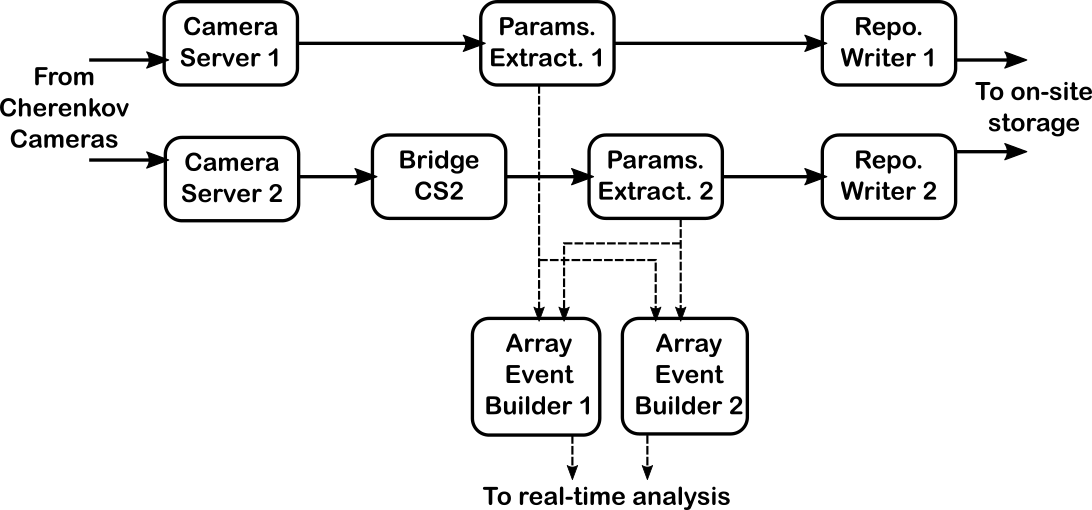}
\caption{Overview of a data acquisition pipeline (DAQ) architecture with two telescopes. Solid lines denote raw event data while the dashed 
lines correspond to event parameters. Camera server 1 has a native interface to the DAQ while camera server 2 has a bridged interface. 
Raw event data are compressed and written to the on-site repository, while extracted event parameters are forwarded to array event builders.}
\label{fig:pipeline_overview}
\end{figure}

The pipeline itself is composed of several modules as seen on figure \ref{fig:pipeline_overview} and it is the evolution of 
the prototyping activities presented at ICRC2015 \cite{modern_middleware}. The \emph{camera server interface} 
allows the pipeline to readout event data from the camera servers. It comes in two flavours: native and bridged. The native 
interface delivers the unified format while the bridged version has a separate component (or bridge) that transcodes the camera 
native format to the unified one. 
The \emph{parameter extraction module} pre-calibrates the events and extracts their high-level parameters, such as Hillas parameters. 
This is done early in the pipeline to help reduce the overall throughput of the system. 
The \emph{array event builder} retrieves event parameters and assembles them into complete array events that are then forwarded 
to the real-time analysis. Due to the high trigger rate of the array (between 30 and 50 kHz), array event building will be load-balanced
across several physical nodes.
The \emph{Repository Writer} receives raw, unified event data from a single telescope, applies a compression algorithm and writes the data
to the on-site repository. The current prototype uses the ZFITS file format \cite{zfits} and custom compression, but the module is flexible enough to allow
other file formats and compressions to be used. \\

Note that the selection of events based on stereo triggers is done entirely by the camera servers and thus not included in this pipeline.\\

The pipeline runs inside the Alma Common Software framework (ACS) \cite{acs} and makes heavy use of ZeroMQ (ZMQ) \cite{zmq} and of the Google 
protocol buffers \cite{protobuf}. It is modular and allows modules to be instantiated as stand-alone executables or ACS component, or 
distributed across several compute nodes. Load-balancing is built-in and thus virtually any data rate can be managed as long as 
the on-site infrastructure can handle it.\\

The pipeline comes with a raw-events reader class written in C++. This reader has been interfaced with \emph{ctapipe} \cite{ctapipe} so that the ZFITS format
can be used readily to analyse the data from early telescopes. 

\section{Camera server readout}

Camera interfacing is done in two different ways: native and bridged. In the native interface the camera teams use the API provided by 
the DAQ pipeline to ship complete events downstream. Provided that the hardware is fast enough, one output stream delivers a throughput
of approximately 9Gbps to a 10Gbps Ethernet interface. The CPU usage is in the order of 1.5 cores. Using infiniband and ip-over-ib instead 
of Ethernet suggests that better performances are to be expected, in the order of 12Gbps per stream. \  
The bridged interface was not investigated very deeply yet, but performances up to 7Gbps for a single stream were achieved while interfacing
Flashcam \cite{flashcam}. This test was done using a dummy Flashcam camera server producing events and sending them downstream using the 
a custom library from the FlashCam project called \emph{TMIO}. The bridge was receiving events via TMIO in the Flashcam native format and was transcoding them to the unified
format in protocol buffers. Unified events were then sent to a ZMQ receiver. 

\subsection{Integration to camera servers}

So far cameras servers used their own code and memory management to build complete single telescope events. Complete events are then copied 
to a protocol buffers data structure before being serialised to a ZMQ socket. This extra copy of the data in memory was avoided by the 
 NectarCam team. Not only did they use the provided API to send events, but they also used it to allocate the memory used to buffer 
the data. This allows incoming event data to be put to the protocol buffer structure with no intermediate copy, thus avoiding this extra copy. This
approach produced the best performances so far: 18Gbps using two parallel streams \cite{nectarcam}. \\

\subsection{LST readout}

\begin{figure}
\begin{centering}
\includegraphics[width=.7\textwidth]{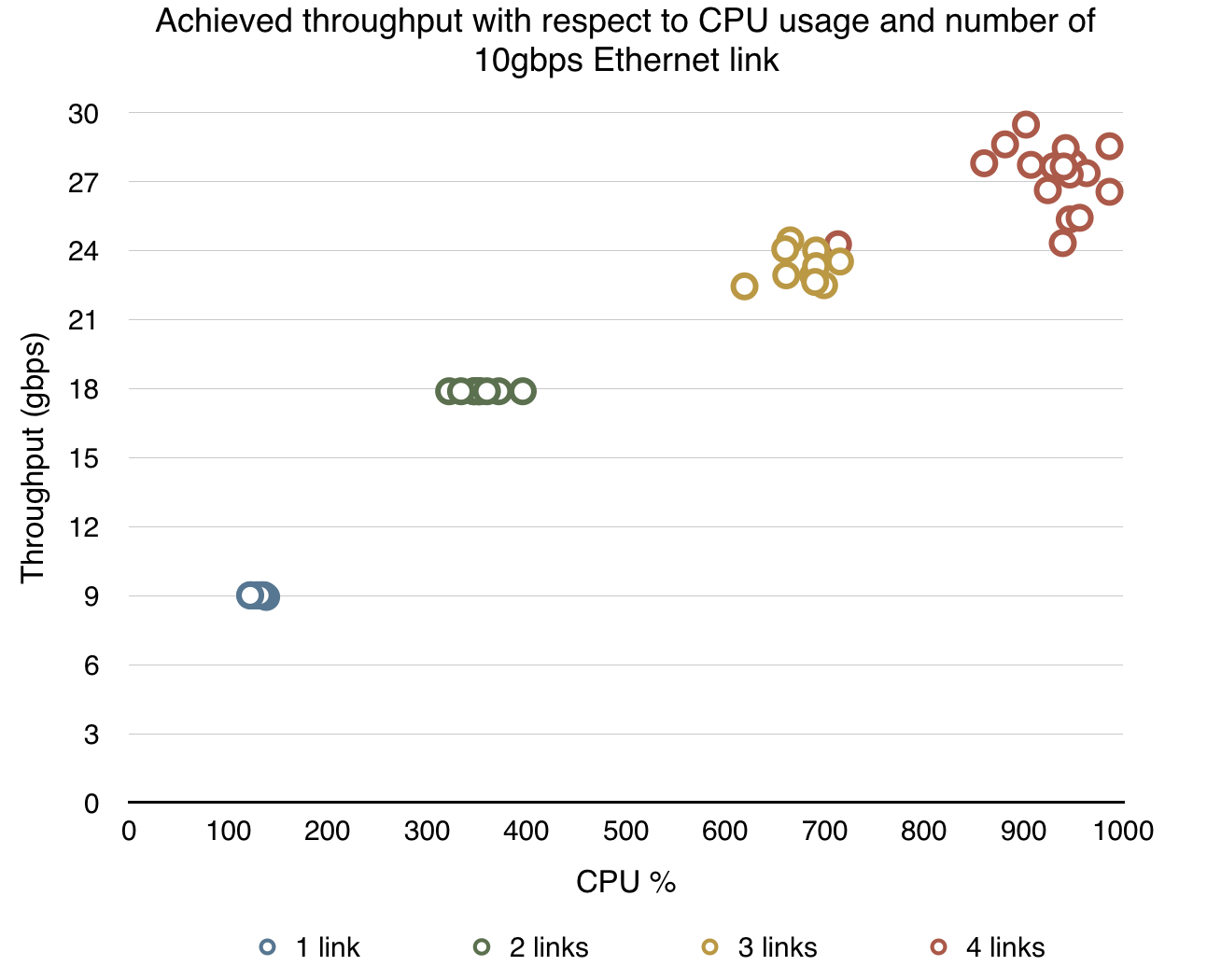}
\caption{Throughput performances using 1, 2, 3 and 4 times 10Gbps interfaces with 1 ZMQ stream per interface. The test was sending 
100 GBytes of event data to remote clients. There was 2x more clients than data producers and the clients just discarded the data. 
The tests was repeated 10 times for 1 to 3 interfaces, and 20 times for 4 interfaces.
}
\label{fig:throughput}
\end{centering}
\end{figure}

It has been recently agreed that the maximum readout data rate for large size telescopes (LST) would be 24Gbps. It is foreseen that 
the LST event builder will reuse the same software as for NectarCam, thus ensuring maximum integration of the interface to DAQ. However, 
the current DAQ software has only been verified up to 18Gbps and pushing the performances up to 24Gbps might be challenging. Indeed, as
shown on figure \ref{fig:throughput}, preliminary tests done with several 10Gbps interfaces indicate that the performances do not grow
linearly if more interfaces / streams are added. This is more likely due to a suboptimal usage of the available resources, for example
if non uniform memory access boundaries are crossed. The 24Gbps throughput from the requirement seems in our reach, however it may need 4 interfaces rather 
than 3 unless further optimisations are made in the software. This will be closely investigated in the coming months, and the most 
simple solution will be selected for the commissioning of LST1.

\section{Processing facilities}
The DAQ pipeline main task is to collect raw event data from the camera servers and to write it to the on-site repository. Besides
this task, it will also run the parameter extraction and array-event reconstruction from these parameters. 

\subsection{Parameter extraction}

The parameter extraction is located early in the pipeline to help reduce the overall throughput. Indeed, event parameters are much 
smaller in size compared to raw events and can be extracted for a single telescope. Sending event parameters 
to the array event builder rather than raw data makes the extra throughput negligible compared to the bulk data transfer. \\
The algorithms used to pre-calibrate the event data will come from the camera teams as they are the ones who know best how to calibrate
their instrument. This algorithm will be reversible so that a better calibration can be applied by the offline analysis pipeline. 
The algorithms used to extract the event parameters will be common to all cameras and most likely run in python inside 
the ctapipe framework. \\
As of now, it remains unclear what will be the details of the interface between the DAQ pipeline and ctapipe. Our first choice is to use
 the protocol buffers and ZMQ once again to interface the DAQ component with a generic ctapipe component running in ACS. 
If better performances are needed, then the ctapipe routines will be called directly from the DAQ C++ component. However, as this approach
is less flexible than our first choice, it will be implemented only if needed. 

\subsection{Array event builder}

The array event builder is the interface between the DAQ pipeline and the real-time analysis. It collects single telescope image parameters
and assembles them into array events based on the stereo trigger information. Due to the high trigger rate of CTA it is foreseen that
more than one instance will be needed. \\

As there might be more than one parameter extraction instances in case of high throughput telescopes (e.g. LSTs), it is not trivial how
to route each event parameter to its array building instance. Early prototyping activities made the array builder request a given event ID
to all parameter extraction nodes. Only the node that had processed this event would in-turn forward the data to the requesting event builder.\\

This approach was successful during prototyping, but it was not implemented up to the scale of CTA. As soon as the final data rates for 
all telescopes will be known, along with the performances of the parameter extraction algorithms, then we will repeat the prototyping
with a more realistic setup.

\section{Interface to the on-site repository}

The interface to the on-site repository is two-fold: read and write. The writing is done by a \emph{Repository Writer} ACS component while 
the reading is done by interfacing a C++ class to the ctapipe framework. 

\subsection{Repository Writer}

\begin{figure}
\begin{centering}
\includegraphics[width=.99\textwidth]{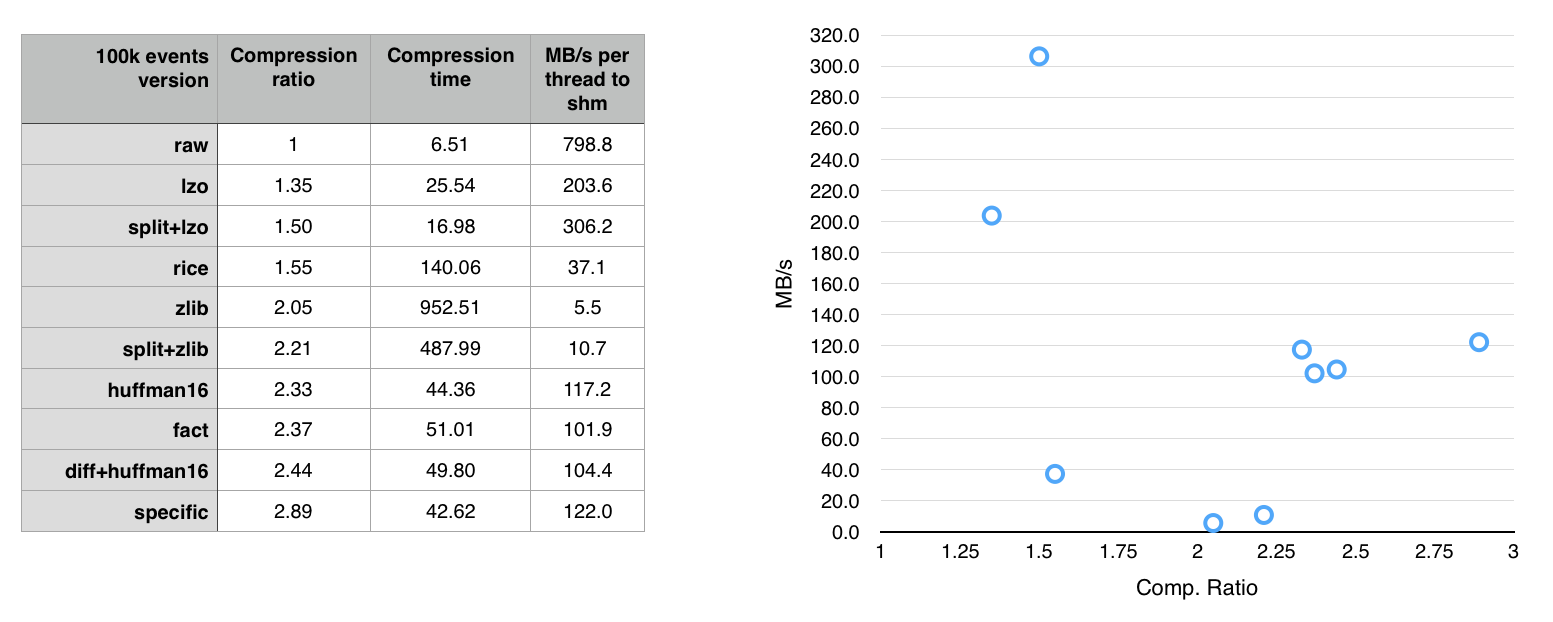}
\caption{Compression ratios and throughput per core for various compression schemes. \emph{raw} means that no compression was applied. \emph{lzo} 
is the well-known real-time compression algorithm \cite{lzo}. \emph{split} means that the data are pre-processed by splitting the high and 
low part of the binary words. rice is the rice compression \cite{rice}. \emph{zlib} uses the well-known gzip algorithm \cite{gzip}. \emph{huffman16}
is an implementation of the huffman coding on 16 bits \cite{huffman}. \emph{fact} is the scheme used by the FACT project to store their data \cite{fact}.
\emph{diff} means that the difference between each sample is stored rather than the samples themselves. \emph{specific} means that 
the best performing algorithm on each type of data (indices, samples, \ldots) was used for each column of the ZFITS file. 
}
\label{fig:comp_ratios}
\end{centering}
\end{figure}

We implemented the repository writer as an ACS C++ component. It listens to ZMQ streams, sorts incoming events and writes them 
in ZFITS format. The compression used in ZFITS is a custom scheme, yet to be finalised once real CTA data will be available. Various
compression schemes were tried out so far, as depicted on figure \ref{fig:comp_ratios}. It seems obvious that a compromise has to be made
between speed and compression ratio. Considering the long-term storage foreseen for CTA data it looks to us that it would be worth applying a 
specific compression algorithm per kind of data stored, as ZFITS allows to do. 

\subsection{ctapipe reader}

\begin{figure}
\begin{centering}
\includegraphics[width=.99\textwidth]{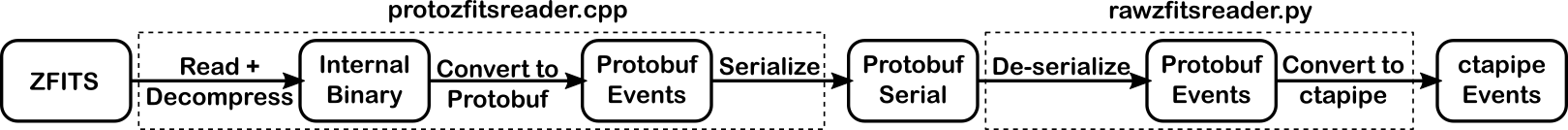}
\caption{ Architecture of the ZFITS reader prototype for ctapipe. The DAQ C++ class ProtoZFitsReader is used to perform the low 
level operations and obtain the events in a protocol buffer object. This object is then serialised and passed to  
the python side of the interface. The serial data are then de-serialised using the protocol buffers for python and converted 
to the data structure expected by ctapipe. This architecture has been easy to implement as only protocol buffer serial data
are passed between C++ and python. It is obviously suboptimal as extra format conversions take place. However, because most of the 
computation time is spent decompressing the data, the overhead remains small, in the order of 3 percent of the total time. 
}
\label{fig:ctapipereader}
\end{centering}
\end{figure}

A prototype implementation of a ZFITS reader for ctapipe was implemented. To avoid extra developments, it reuses the C++ class from 
the DAQ pipeline instead of reading the data natively in python. The data are read 
and uncompressed by the C++ code, serialised in protocol buffer format and given to python as a binary block. 
A lightweight python layer then decodes this binary block and delivers it to the expected ctapipe structure. 
A detailed view of the architecture of the reader can be seen on figure \ref{fig:ctapipereader}. \\

This approach had the advantage that it has been fast and effective to implement, thus allowing immediate usage by the early prototypes. 
The reader could be made more efficient if the interface between the C++ class and python would be made natively rather than via the 
protocol buffers. Even though the overhead is very small, if ZFITS is selected as CTA raw data format then we will consider to make 
the effort to improve this interface. 

\section{Future work}

The prototyping activities to interface remaining Cherenkov cameras is ongoing. As soon as the LST event builder becomes available we will
perform the benchmark required to understand if the proposed architecture could work for such data rates. \\

In parallel, a prototype interface to the real-time analysis is being developed. We are currently focused in embedding the parameter
extraction into the DAQ pipeline and will soon move to the updated version of the array event builder.\\

Once the data format for the raw data of CTA will be selected, we will make the required adjustments to the Repository Writer component.

\section{Acknowledgments}
This work was conducted in the context of the ACTL working group of the CTA Consortium. We gratefully acknowledge financial support 
from the agencies and organisations listed here: http://www.cta-observatory.org/consortium\_acknowledgments

\end{document}